\documentclass[aps,showkeys]{revtex4-1} 

\usepackage{color}
\usepackage{graphicx}
\usepackage{subcaption}
\usepackage{tabularx}
\usepackage[T1]{fontenc}
\usepackage{amsfonts}
\usepackage{amsmath}
\usepackage{amsthm}
\usepackage{rotating}

\newtheorem{definition}{Definition}
\newtheorem{lem}{Theorem}

\begin{document}
\title{Predictive maintenance solution for industrial systems - an unsupervised approach based on log periodic power law}
\author{Bogdan {\L}obodzi\'{n}ski}
\email[]{lobodzinskibogdan@gmail.com}
\affiliation{Burckhardt Compression AG \\
Franz-Burckhardt-Strasse 5, P.O. Box 3352, CH-8404 Winterthur, Switzerland}
\date{\today}

\begin{abstract}
A new unsupervised predictive maintenance analysis method based on the renormalization group approach used to discover critical behavior in complex systems has been proposed. 
The algorithm analyzes univariate time series and detects critical points based on a newly proposed theorem that identifies critical points using a Log Periodic Power Law function fit. 
Application of a new algorithm for predictive maintenance analysis of industrial data collected from reciprocating compressor systems is presented. 
Based on the knowledge of the dynamics of the analyzed compressor system, the proposed algorithm predicts valve and piston rod seal failures well in advance.
\end{abstract}


\maketitle
\nopagebreak

\section{Introduction}
\label{section:introduction}

Detecting the symptoms of a failure and predicting when it will occur, in multivariate or univariate time series collected via the Internet of Things (IoT), is central to the concept of predictive maintenance (PM), which is now used in almost every area of industry.
PM allows a company to better prepare for a potential failure by redesigning the production process in advance or creating a workaround when shutdown is not possible. Thus minimizing the costs and the effort of standard maintenance operations through predictive engineering.

Predicting failures, provided by PM, can be very profitable for a company, under the condition that PM minimizes the number of false warnings (false positives) and maximizes the number of correctly predicted events (true positives). 

Let us define failure as the termination of the ability of a part of an industrial device to perform the required function with the required efficiency. Another aspect of a failure that needs to be specified is how it can manifest itself.
A failure can develop gradually or appear as a sudden event. In this work we consider only those failures that develop over time. 

Creating a properly working PM process faces two main problems:
\begin{enumerate}
\item related to the determination of time when a failure occurred in the considered technological process or IoT data,
\item the development or the application of the best algorithm 
(based on physical description, machine learning, or statistical methods) to the data being analyzed. 
\end{enumerate}
Proper identification of the failure in IoT data (data labeling process) is a very difficult task and usually requires very specialized knowledge. 
For various reasons (economic, technological), not every failure requires corrective action. 
This is because the required efficiency that defines a failure in the monitored unit may depend on the degree of reliance of the process on the considered failure.
Sometimes, a minor failure is not a sufficient reason to stop a monitored machinery or a production process. 
The determination of so-called good IoT data, i.e., the data when the monitored device behaves normally, is also a data labeling process that requires expertise.  
In this work, every action that defines the state of the data is considered a data labeling process.

Attempts to build a solution based on supervised Machine Learning (ML) models will encounter the following problems:
\begin{enumerate}
\item the complexity of the data labeling process,
\item the model degradation: due to the evolution of IoT data, it is necessary to update the supervised model because the current one starts to behave incorrectly. Such an update forces the creation of new training data necessary for a new supervised model (new data labeling process),
\item the complex process of monitoring the predictions of the currently working model. This is a difficult task, especially when dealing with a small number of predicted failures \cite{monitoring0}. 
\end{enumerate}
In published works on failure prediction using supervised solutions \cite{review0}, the authors omit the need for a continuous data labeling process, as well as the need to monitor the degradation of the running model.
Thus, an important economic part of PM projects based on supervised models is passed over in silence.
In review articles \cite{review0}, methods based on Neural Network architecture are placed in the category of unsupervised methods. 
However, these methods also require labeled training data. In this article, the term “unsupervised method” is understood only as a solution that does not require any knowledge of labeled data and any information about the state of the input data as good/healthy or anomalous. 

The complex and expensive cycle of building and implementing a supervised method for a predictive maintenance process, sketched above, strongly suggests that unsupervised approaches are a better tool for building PM processes.
An unsupervised solution removes the necessity of: 
\begin{enumerate}
\item a complex data labeling process running at all times as a part of a supervised model,
\item monitoring of the deterioration of the model quality and thus re-creating models by training them on updated training data.
\end{enumerate}

This paper describes an unsupervised failure prediction method \cite{patent0} used to monitor reciprocating compressor systems based on a concept called Log-Periodic Power Law (LPPL) proposed in \cite{Johansen2000} and \cite{Sornette2003ab}.
The LPPL mechanism is used to predict disasters in financial data \cite{finance0}, earthquakes \cite{earthquakes0}, avalanches and similar phenomena \cite{avalanches0}.

However, due to the different nature of the data analyzed, the LPPL method cannot be applied in the same way as for 
bubble (or anti-bubble) detection in economic time series. Due to the different definitions of PM failure in industrial applications, modifications need to be made to the LPPL method. 
In a case of data describing the financial market, the variable that directly characterizes changes in the system under analysis is examined.  
Such a variable is, for example, the index of the analyzed financial market or the prices of shares.

In the case of a machine (e.g. compressor or other systems containing a number of subsystems), the analysis uses {\it indirect} data collected by sensors. 
It is not possible to measure the direct changes causing the failure (e.g. material degradation, cracks etc.) 
but it is possible to measure changes resulting from the influence of deteriorating machine components on the measured values.

In the description of the application of the presented method, such indirect data are, for instance, changes in the opening angle of suction or discharge valves inside the compression chamber as a function of the volume in the cylinder chamber expressed by the angle of rotation of the crankshaft.   
In other words, degradative (unmonitored) changes affect measured variable changes in a less visible (more distorted) ways than in case of a direct measurement of variables.

The principle of operation of the proposed method is based on the detection of such behavior of the analyzed data, which is in accordance with the logarithmic periodic power law. 
Finding such a functional behavior in the data is equivalent to finding the critical point of the system. It corresponds to the situation in which the system begins to work with decreasing efficiency, which is a signature of future failure. 
The point in the time series thus identified defines the start of the countdown to the predicted failure.
Let's call the trend change point {\it the initial breakdown} (IB).
The concept of detecting an IB point for use in predicting failure is similar to the concept of detecting a trend change point in a time series.  
Therefore, it is not necessary to predict the time of failure in the future. 
All that is needed is to determine whether a given point in the time series is or is not an IB starting point.
If it is known that the current point in the analyzed time series is an IB point, then based on the knowledge of the dynamics of the monitored device, it is possible to determine the time window in the future (in units of the device's operating time) in which the failure will occur.\\
It should be emphasized that the determination of the IB point is not equivalent to the detection of anomalies. Anomalies in the behavior of the unit should be expected only in the time window of the expected failure.  
The concept of a time window in the future during which a failure is predicted is due to the fact that it is not possible to predict the exact time of failure of the monitored device using the presented method.
This is mainly due to unpredictable events, such as: changes in the load of the device, periods when the device is off or independent maintenance activities.

The organization of the paper is as follows.
Chapter \ref{section:related_works} briefly describes the current status of unsupervised predictive maintenance in the industry.
Chapter \ref{section:lppl} answers the question of why the logarithmic-periodic oscillation detection method is suitable for detecting sudden failures in an industrial system.
Chapter \ref{section:fitting} describes a numerical method for determining failure time points from the data. 
The part \ref{section:datadescription} introduces the data used for the analysis presented and the following
chapter \ref{section:predictions} shows the results obtained using the new method and compares these results with those obtained using online statistical method.  
Finally, the part \ref{section:discussion} discusses the results and summarizes the advantages and disadvantages of the proposed method. 
This section also briefly outlines the criterion for selecting data to be used by this method. 

\section{Related works} 
\label{section:related_works}

Given the introduced definition of “unsupervised method”, as a methodology that does not require knowledge of the state of the input data, the list of existing methods for failure prediction is very limited.
Following the work of \cite{survey_1} (which also includes references to other research works, as well as studies of specific solutions involving a broader understanding of the "unsupervised method"), the spectrum of available solutions can be generally divided into 2 categories. 
\begin{enumerate}
\item Solutions involving the use of prediction techniques to predict points in the future and then compare them with actual data to detect anomalies, as shown in \cite{method_1}. 
Depending on predictive solutions used, this category of methods requires a large amount of data.
In this case, a separate problem is the accuracy of the prediction part. Where the prediction part is a supervised method. 
Hence the requirement to control it, which makes the whole solution complicated if one wants to use it in practical applications.
\item Solutions based on measurements of distance or similarity used to evaluate the degree of data anomaly.
In this case, methods using clustering algorithms \cite{method_2} or determining similarity between time series or data are used.
This approach sometimes leads to problems when encountering new data that did not exist in the past. 
It happens that it makes it impossible to use them for real-time data analysis.
\end{enumerate}
The solution proposed in this paper is an attempt to solve the issues identified in both of the above categories:  
the problem of demand for large amounts of data and the problem of data changes that do not occured in the past.

The essence of the proposed method is to determine the initial breakdown (IB) points in the time series of input data. Thus, it is natural to compare the proposed model with online methods that determine trend change points in time series in fully unspervised mode. 
Latest advances in statistical online trend change point detection can be found in studies \cite{changepoint1,changepoint2}. 
The difference between the statistical methods of determining the trend change points and the IB points determined by the proposed method is that in the proposed solution we are based on the search for a specific functional behavior (pattern) in the data. 
As will be shown later, this pattern is characteristic for physical processes leading to a degradation 
of the monitored components of the device
A comparison of the results of the two methods, based on the same input data (statistical \cite{changepoint2,changepoint3} and the proposed model), will be presented and discussed in the section \ref{section:predictions}.

\section{The occurrence of log-periodic oscillations as a prelude to failure.}
\label{section:lppl}

The purpose of the PM method is to provide predictions for future failures of the system described as a set of various components cooperating together.  
This section illustrates why the LPPL-based algorithm is applicable to failure prediction and describes the basics of the LPPL-based method.

The generalized relation describing the hazard rate $h\left(t\right)$ (or hazard function) of a certain physical quantity at time $t$ prior to material destruction by degeneration \cite{Voight1988} is of the form
\begin{equation}\label{basic_law_1}
\dot{h}\left(t\right) = G h\left(t\right)^{\delta}
\end{equation}
\noindent
where the $\dot{h}\left(t\right)$ denotes the derivative of the function $h\left(t\right)$ in time $t$. The $\delta$ and $G$ are the parameters of model. The $G$ is a positive constant parameter.
In the following, it is assumed that $h\left(t\right)$ corresponds to changes in the variable that is tracked and from which the failure is attempted to be predicted. 
The hazard function, based on conditional probability theory, measures the probability that the relevant variable will show signs of a critical behaviour, given that the critical behaviour has not occurred prior to time $t$.

The equation (\ref{basic_law_1}) has 3 classes of solutions depending on the value of $\delta$. 
For $\delta=1$ the exponential function is obtained
\begin{equation}\label{solution_exp}
h\left(t \right) = h\left(t_{0}\right) e^{G\left(t - t_{0}\right)}
\end{equation}
\noindent
where $t_{0}$ is a value of the initial time.\\
For $\delta < 1$:
\begin{equation}\label{solution_1}
h\left(t \right) = \left[\left(1-\delta\right) \left(t - t_{0}\right) G \right]^{\frac{1}{1-\delta}}
\end{equation}
\noindent
and for $\delta > 1$:
\begin{equation}\label{solution_2}
h\left(t \right) = \left[\left(\delta-1\right) \left(t_{c} - t\right) G \right]^{\frac{1}{1-\delta}}
\end{equation}
\noindent
with $t_{c}$ as a constant corresponding to time in a future.
As can be seen, the solutions found for $\delta = 1$ (\ref{solution_exp}) and for $\delta < 1$ (\ref{solution_1}) do not converge for times $t > 0$. 
Therefore, the time of critical behaviour cannot be determined. 
The most interesting case is $\delta > 1$ (\ref{solution_2}), 
where the solution has a converging point at a finite time $t_{c}$ in future. 
In our analysis, the time $t_{c}$ will denote a critical time point and can be determined by a choice of 
$G=\left(\frac{1}{t_{c}}\right)^{\frac{1}{1-\delta}}$. Then $t_{c}=\frac{1}{\left(\delta-1\right) h\left(t_{0}\right)^{\delta-1}}$. \\

Let us assume that the degradation of a working part, can be treated as a discontinuous stochastic process associated with a given monitored variable. 
To simplify the analysis, the second assumption is to treat the degradation changes of physical quantity $p\left(t\right)$ over time $t$ as a non-homogeneous Poisson process in which the changes occur according to the hazard rate function $h\left(t \right)$. 
The dynamics of such a process can be described by the equation 
\begin{equation}\label{Degradation_process_base}
d p\left(t\right) = - p\left(t\right) h\left(t\right) dt
\end{equation}
\noindent
the solution of which can be written as
\begin{equation}\label{Degradation_process}
\log{\left[\frac{p\left(t\right)}{p\left(t_{0}\right)}\right]} = - \int_{t_{0}}^{t} h\left(u \right) du = P\left(t\right).
\end{equation}
\noindent
In our case, $P\left(t\right)$ has an approximate form 
\begin{equation}\label{Degradation_process_solution}
P\left(t\right) \approx \frac{h_{0}}{\eta + 1} \left(t_{c} - t\right)^{\eta+1}
\end{equation}
\noindent
where $\eta = \frac{1}{1-\delta}$ and the $P\left(t\right)$ is shifted by the integration constant $\frac{h_{0}}{\eta + 1} \left(t_{c}-t_{0}\right)^{\eta+1}$.
The result obtained (\ref{Degradation_process_solution}) coincides with work \cite{Ledoit2000} (compare with equation (3) in the reference \cite{Ledoit2000})

Solution (\ref{Degradation_process_solution}) is invariant under continuous scale invariance (CSI), which manifests itself through the scaling property of the solution $P\left(t \right)$ if the argument of the function $P\left(t \right)$ is scaled to ($t_{c} - t$).   
Rescaling by some factor $\nu$ the argument $t_{c} - t \rightarrow \left(t_{c} - t\right)\times\nu$ changes the solution 
$P\left(t \right)$ to the form $P\left(t \right)\times\mu$ where $\mu = \nu^{-1-\eta}$ \cite{IdeSornette2002}.
The CSI feature, around the critical points $t=t_{c}$, is common to systems demonstrating a continuous phase transition (second order phase transition).

The basic assumption of the LPPL method is that, the described process is near the critical point of the second-order phase transition.
With this assumption, the final equation is obtained, the use of which for fitting is known in literature as the LPPL method 
\cite{Feigenbaum1996, Sornette1996}.

Let $W\left(t\right) = \log\left(p\left(t\right)\right)$ and $p\left(t\right)$ corresponds to the variable by which our industrial system is analyzed as a function of time $t$.
Let $t_{c}$ be the time of event that defines our phase transition in physical framework (critical time point). 
Then the argument $x$ and the real function $F\left(x\right)$ are defined as
\begin{equation}\label{notation1}
x = t_{c} - t \textrm{ with } t < t_{c} \textrm{ and } F\left(x\right) = W\left(t_{c}\right) - W\left(t\right)
\end{equation}
\noindent
As a starting point for our derivation, the CSI is assumed to exist around critical points.

This allows us to use the renormalization group approach, which permits us to write a certain real function $F\left(x\right)$ around a critical point $x\approx 0$ through the rescaled argument $x$ expressed by a scaling function $\phi\left(x\right)$ in the form 

\begin{equation}\label{scaling1}
F\left(x\right) = \frac{1}{\mu} F\left[\phi\left(x\right)\right]
\end{equation}
\noindent
where $\mu$ is a constant and its argument $x$ is invariant under arbitrary linear transformation of the $x$:
\begin{equation}\label{scaling2}
\phi\left(x\right) = \nu x \textrm{ and } x>0
\end{equation}
with $\nu$ as a constant.  

\noindent
The solution of equation (\ref{scaling1}) is the function
\begin{equation}\label{solution0}
F\left(x\right) = C x^{\alpha}
\end{equation}
\noindent
where $C$ and $\alpha$ are constants to be determined. 

The condition of scaling invariance (\ref{scaling1}) with the general form of the solution postulated 
by eq. (\ref{solution0}) can be rewritten as
\begin{equation}\label{deriv1}
C x^{\alpha} = C \frac{\nu^{\alpha}}{\mu} x^{\alpha} 
\end{equation}
\noindent
which, after taking into account the identity,
\begin{equation}\label{identity1}
1 = e^{i 2 \pi n} \mbox{ with } n \in \mathbb{N}, 
\end{equation} 
leads to the equation
\begin{equation}\label{deriv1a}
e^{i 2 \pi n} = \frac{\nu^{\alpha}}{\mu}
\end{equation}
what allows to calculate the $\alpha$ exponent in a most general form as
\begin{equation}\label{deriv2}
\alpha  = \frac{\log\left(\mu\right)}{\log\left(\nu\right)} + i \frac{2 \pi}{\log\left(\nu\right)} n .
\end{equation}
\noindent
Therefore, the solution of equation (\ref{scaling1}) can be expressed, according to \cite{Nauenberg1975}, as 
\begin{equation}\label{deriv3}
F\left(x\right) = \frac{C}{\mu} \left(\nu x\right)^{\frac{\log\left(\mu\right)}{\log\left(\nu\right)} + i \frac{2 \pi}{\log\left(\nu\right)} n} = 
\frac{C}{\mu} \left(\nu x\right)^{\frac{\log\left(\mu\right)}{\log\left(\nu\right)} } \Pi\left(
\frac{\log\left(x\right)}{\log\left(\nu\right)}
\right)
\end{equation}
\noindent
where $\Pi\left(\cdot\right)$ is a periodic function with period $1$, i.e. $\Pi\left(y\right) = \Pi\left(y + 1\right)$ .
\noindent
The index $n$ should be treated as one of the parameters characterizing the described physical system. 
Since $n$ and other parameters appearing in the function (\ref{deriv3}) are unknown, 
it is necessary to reformulate the function (\ref{deriv3}) in such a way that it can be used to fit existing data 
and thus determine whether a given point $t_{c}$ is a critical point.

An additional necessary condition that must be satisfied by function (\ref{deriv3}), or more precisely
by its real part, is the trend that is determined by the power law ($n=0$), which is the leading order term
and the oscillations associated with $n\neq 0$ will contribute as next-to-leading order corrections.

The periodic function $\Pi\left(\cdot\right)$ can be expressed by means of the Fourier series with respect to the variable $y$ with period $T$ 

\begin{equation}\label{fourier1}
\Pi\left(y\right) = \exp\left[ i 2 \pi n \left(\frac{y}{T}\right)\right] = \sum_{k=-\infty}^{+\infty} c_{k} e^{i 2 \pi k \left(\frac{y}{T}\right)}
\end{equation}
\noindent
with 
\begin{equation}\label{fourier2}
c_{k} = \frac{1}{T} \int_{-\frac{T}{2}}^{\frac{T}{2}} \left\{
\exp\left[ i \frac{2 \pi}{T} y n\right]
\right\}
e^{-i\frac{2 \pi}{P} k y} dy = \frac{1}{\pi}\frac{\sin\left[\pi\left(k - n\right)\right]}{k-n}  \mbox{ for } n, k \in \mathbb{N} .
\end{equation}
\noindent
However, non-zero coefficients $c_{k}$ of the Fourier expansion (\ref{fourier1}) are obtained only for non-integer differences 
$\left(k-n\right)$, which contradicts our claim for the existence of non-zero expressions associated with $n \in \mathbb{N}$.
The problem of zero coefficients of the fourier expansion of (\ref{fourier1}) can be solved by rewriting the identity (\ref{identity1}) to the form 
\begin{equation}\label{identity2}
1 = e^{i 2 \pi n} \rightarrow 1 = e^{i \frac{2\pi}{q} \left(n q\right)}
\end{equation}
\noindent
where, the variable $q$ denotes a parameter associated with the physical degradation mechanism of the described system.

In that formulation, $\alpha$ (\ref{deriv2}) can be rewritten as
\begin{equation}\label{deriv2a}
\alpha  = \frac{\log\left(\mu\right)}{\log\left(\nu\right)} + i \frac{2 \pi}{\frac{\log\left(\nu\right)}{q}} \frac{n}{q}  
\end{equation}
\noindent
what allows us to rewrite the equation (\ref{deriv3}) to the form
\begin{equation}\label{deriv3a}
F\left(x\right) = \frac{C}{\mu} \left(\nu x\right)^{\frac{\log\left(\mu\right)}{\log\left(\nu\right)} + 
i \frac{2 \pi}{\frac{\log\left(\nu\right)}{q}} \frac{n}{q} }
\end{equation}
\noindent
In this case, the expansion of $F\left(x\right)$ into Fourier series gives us the following result.

\begin{equation}\label{deriv4}
F\left(x\right) =
\frac{C}{\mu} 
\left(\nu x\right)^{\frac{\log\left(\mu\right)}{\log\left(\nu\right)}}
e^{i \frac{2 \pi}{\frac{\log\left(\nu\right)}{q}} \frac{n}{q} \log\left(x\right) } 
= 
\frac{C}{\mu} 
\left(\nu x\right)^{\frac{\log\left(\mu\right)}{\log\left(\nu\right)}}
\sum_{k \in \mathbb{N}} 
c_{k} e^{i \frac{2 \pi}{\frac{\log\left(\nu\right)}{q}} \log\left(x\right) k} 
\end{equation}
\noindent
where
\begin{equation}\label{coeff1}
c_{k} = \frac{1}{T} \int_{-T/2}^{T/2} e^{i\frac{2 \pi}{T}\left(\frac{n}{q}-k\right)y} dy = 
\frac{1}{\pi} \frac{\sin\left[\pi \left(k - \frac{n}{q}\right)\right]}{ k - \frac{n}{q}}
\end{equation}
\noindent
with a redefined variable $y=\frac{\log\left(x\right)}{ \frac{\log\left(\nu\right)}{q} }$.

Given the denominator $\left(k - \frac{n}{q}\right)$ in the coefficients of $c_{k}$ (\ref{coeff1}), 
the main dominant terms of the series (\ref{deriv4}) are defined by the index $k$ from the set 
$\left\{-1+\left[n/q\right], \left[n/q\right],1+\left[n/q\right]\right\}$ 
where $[n/q]$ is the integer part of division $n/q$. 
To simplify the notation, let us redefine the index values 
from $\left\{-1 + [n/q] , [n/q] , 1 + [n/q]\right\}$ to $\left\{-1,0,+1\right\}$.

\noindent
This allows us to approximate the final form of the function $F\left(x\right)$ (\ref{deriv4}) 
by the first 3 largest components of the Fourier series ($c_{0}$ and $c_{-1}$ or $c_{+1}$)
\noindent
\begin{equation}\label{deriv6}
F\left(x\right) \approx {} 
\frac{C}{\mu} 
\left(\nu x\right)^{\frac{\log\left(\mu\right)}{\log\left(\nu\right)}}
\left[
c_{0} + 
c_{\pm1}\cos\left( \frac{2 \pi}{\frac{\log\left(\nu\right)}{q}} \log\left(x\right) \right) \pm \right. \left.  i
c_{\pm1} \sin\left( \frac{2 \pi}{\frac{\log\left(\nu\right)}{q}} \log\left(x\right) \right) 
\right]
\end{equation}
\noindent
where the notation $c_{\pm1}$ was used to denote ambiguity as to which coefficient is the second dominant one for $k=-1$ or $k=+1$.
Since only the real part of the expression is of interest (our measurements are real values), 
using the previous definition of the variable $F\left(x\right)$ (\ref{notation1}) and 
generalizing our unknown parameters ($\mu$, $\nu$, $C$, $q$, $n$, $k$ and $W\left(t_{c}\right)$) by adding constant $A$ and phase $\Phi$ to the formula (\ref{deriv6}) to their new representations 
($A$, $B$, $m$, $C$, $\omega$, $\Phi$) one obtains the final formula which is referred to as a first-order model and used in LPPL literature \cite{Feigenbaum1996, Sornette1996}

\begin{equation}\label{final_fit_0}
\begin{split}
W\left(t\right) \approx A + \left|t_{c} - t\right|^{m} 
\left[
B + C \cos\left(\omega \log\left|t_{c}-t\right| + \Phi\right) 
\right] .
\end{split}
\end{equation}
\noindent
For the purpose of numerical fitting the LPPL function to the data, 
a transformed version of the formula (\ref{final_fit_0}) is used, in the form of 
\begin{equation}\label{final_fit_1}
W\left(t\right) \approx 
A + \left|t_{c} - t\right|^{m} 
\left[
B + C_{1} \cos\left(\omega \log\left|t_{c}-t\right|\right) + \right. \left. C_{2} \sin\left(\omega \log\left|t_{c}-t\right|\right)
\right]
\end{equation}
\noindent
where $C_{1} = C\cos\left(\Phi\right)$ and $C_{2} = -C\sin\left(\Phi\right)$.
Both equations (\ref{final_fit_0}, \ref{final_fit_1}) can be used to find critical time points in the time series of input data.
\section{Fitting method of the LPPL model to the data}
\label{section:fitting}

Due to the number of parameters ($A$,$B$,$m$,$C_{1}$,$C_{2}$,$\omega$) necessary to be determined during the fitting procedure of the LPPL function (\ref{final_fit_1}) and the presence of many local extremes, the procedure of obtaining the best fit is difficult and computationally expensive.

Instead of trying to determine the critical time $t_{c}$ in the future, as is determined in the case of predictions of crashes in financial time series \cite{Sornette1996, Shu2019, Ledoit2000}, it is assumed that the critical time point is "now".
With this assumption, the calculations involved in fitting the function (\ref{final_fit_1}) to the data are performed for a range of time windows of different lengths from the past to "now".
This corresponds to the hypothesis that the time point $t_{c}$ (time point of the phase transition) is "now" and corresponds to the last point in the input time series $t_{inp}$.  
Therefore, it is necessary to add to the set of parameters, an additional parameter specifying the length of the subset of time series points $t_{inp}$ preceding the time point $t_{c}$ for which the best fit of the function $W\left(t\right)$ (\ref{final_fit_1}) was found. 
The length of this subset is denoted as  $l_{max}$. \\
In particular, our redefined set of arguments $x=t_{c}-t$ in the matching procedure is the set of points 
$\left< 1, x_{l_{max}} \right>$ 
in time units characteristic to the $t_{inp}$ series.
The definition of the argument set $x=t_{c}-t$ in the matching procedure is replaced by the set of points 
$\left<1, x_{l_{max}}\right>$, where $1$ corresponds to the present time and $l_{max}$ corresponds to the time of $l_{max}$ from the past.
The index $l_{max}$ is the one of the parameters of the matching function (\ref{final_fit_1}).

As parameter fitting, the method described in the work of \cite{Shu2019} is used,  
appropriately modified for our purposes, i.e., by excluding the parameter corresponding to the $t_{c}$  
and adding the parameter $l_{max}$ to the fitting procedure. 

The constraints imposed on the fitting parameters ($l_{max}$,$A$,$B$,$m$,$C_{1}$,$C_{2}$,$\omega$) are as follows:
\begin{enumerate}
\item $l_{max}$: the number of past data used for the best fit. For a small number of data there may be too many good fits (with very small fitting error), which may correspond to random correlations of the data with the form of the fitted function (\ref{final_fit_1}).
\item $A > 0$: since in our case there are always positive values. It is determined by the character of the input data. 
\item $0<m<1$: to ensure that the fitting value for the critical time $t_{c}$ was greater than zero ($m>0$) and changed faster than exponentially for times close to the critical time $t_{c}$ ($m< 1$). 
\item $2<\omega<8$: 
this condition avoids too fast log-period oscillations (otherwise they would fit the random component of the input data) and too slow log-period oscillations (otherwise they would contribute to the power law behaviour $\approx A + B \left|t_{c} - t\right|^{m}$)
\item $B$, $C_{1}$,$C_{2}$: these parameters are fitted without additional constraints.
\end{enumerate}
Depending on the type of input data to be analyzed, the limits of variation of the parameters to be fitted require careful adjustment.

Having determined all parameters of the fitting LPPL function, 
in the next step, it is necessary to determine trends based on the identified local maxima and minima of the shape of the fitted function: $T_{max}$ for local maxima and $T_{min}$ for local minima. 
The procedure of finding trends is carried out separately for maxima and minima in the following way:
\begin{enumerate}
\item all local extrema (N) are found,
\item from this set, the N-1 extreme values closest to the current time point (i.e., the point for which an attempt is made to determine whether phase transition has occurred or not) are selected,
\item To this set of points a straight line is fitted by linear regression. The slope of the line determines the trend for a given category of extremes (maxima or minima).
\end{enumerate}

Then, using the calculated trends of the extreme values of the best LPPL fit, it is determined whether a given point 
(defined as a pair: $\{$datetime, value$\}$) corresponds to a phase transition or not. 
For this purpose, a theorem describing the critical time point was formulated.
Its proof will be the aim of the next publication.

\begin{lem}[]\label{bl:theorem1}
Assume that the function $f\left(x\right)$ corresponds to the best found fit of the LPPL function (\ref{final_fit_1}) to the analyzed input time series $ts=\{\left(t_{n}-t_{l_{max}},y_{n-l_{max}}\right),...,\left(t_{n}-t_{n-1},y_{n-1}\right)\}$ where $l_{max}>0$ is one of the fit parameters of the function $f\left(x\right)$ (\ref{final_fit_1}) 
specifying the length of the sequence preceding the actual values $\left(t_{n}-t_{n-1},y_{n-1}\right)$ defined by the index $n$.
The function $f\left(x\right)$, have $N_{max}>2$ local maxima and $N_{min}>2$ local minima. 
Let $T_{max}$ denote the slope of the linear fit determined by the last $N_{max}$ values of the local maxima and 
$T_{min}$ denote the slope of the linear fit determined by the last $N_{min}$ values of local minima.

If both trends $T_{max}$ and $T_{min}$ determined for the function $f\left(x\right)$ have the same behavior (both increasing or decreasing) for the last point $\left(t_{n}-t_{n-1},y_{n-1}\right)$ of the series $ts$, then the point $\left(t_{n},y_{n}\right)$ is the critical point for the series $ts$. The trend of the series $ts$ will change to the opposite for next points $\left(t_{n+k}, y_{n+k}\right)$ (where $k>0$) with respect to the trend for points preceding $t_{n}$. 
\end{lem}

In the presented method, any point that satisfies the Theorem (\ref{bl:theorem1}) is treated as a point initiating a future failure - the initial breakdown (IB) point. 

The formalism described above can be summarized as follows:
the appearance of a failure due to deterioration of components in a complex system is preceded by the appearance of a phase transition of the 2nd kind, defined by point IB (critical time point). From this time point, the process of failure development begins. 
This period (after the time defined by the IB point) can be monitored by detecting anomalies. Anomalies (as a result of reduced process efficiency) occur when the effects of the failure become large enough and begin to affect the behavior of the monitored device. 

Determining the time of failure requires knowledge of the dynamics of the monitored system and will be discussed in section \ref{section:predictions}.

\section{Data description}
\label{section:datadescription}

Monitoring data from a reciprocating compressor, describing the PV diagram of one of the compression chambers, was used to demonstrate the operation of the method. 
Based on these, the values of the angle of opening of the suction valve (OSV) expressed by the angle of rotation of the crankshaft were determined.
The data is available at \cite{data_private}.

This value, for a given cycle described by the PV diagram, is very sensitive to changes in the amount of gas in the chamber, for example, due to its leakage through a broken valve or piston rod seal system. OSV changes are directly dictated by the thermodynamics of the compression process in the compressor chamber.  
Monitoring the changes in OSV provides a base for compressor diagnostics and allows to determine compressor efficiency, valve operation, or the condition of the piston seals or piston rod sealing elements \cite{Reciprocating_compressors}.

The data has been averaged to daily values and covers a period of time between 2019-08-23 and 2022-01-19.

In order to compare the results of failure prediction, data identifying the dates of repair interventions with their respective reasons for failure and the dates of observation of anomalous compressor behavior without interruption of operation were used.  

\section{Prediction of failures: methodology and results}
\label{section:predictions}

To test the effectiveness of our algorithm, a backtest of the detection method was conducted on the OSV data, calculating initial breakdown (IB) points. 
The range of the variable length of the time series $l_{max}$ was assumed to be $ 30< l_{max} <101$ in time units of days.
Given the minimum number of observations (101 days) needed to perform calculations of the IB points, the backtest is started for time $t = t_{start} + 101$ (in daily units). Then, moving forward in time to the future, the best-fit LPPL function (\ref{final_fit_1}) is calculated for each subsequent time $t$ by determining the goodness of fit of the LPPL function using the mean squared error ($mse$).

Intuitively, one can expect that the accuracy of determining the critical points using Theorem (\ref{bl:theorem1}) will strongly depend on the error of fitting the LPPL curve (\ref{final_fit_1}) to the data.  
The smaller the fitting error, the greater the confidence that a given point of the input time series is indeed an IB point according to Theorem (\ref{bl:theorem1}).
In addition, it is expected that in the vicinity of the true IB breakpoint (before and after it), the method should find some good fits of the LPPL function with a small error, but this is not a necessary criterion for the existence of an IB point for days as time units.

\begin{figure}[h!]
    \centering
    \begin{subfigure}{0.9\textwidth}
        \includegraphics[width=1\linewidth]{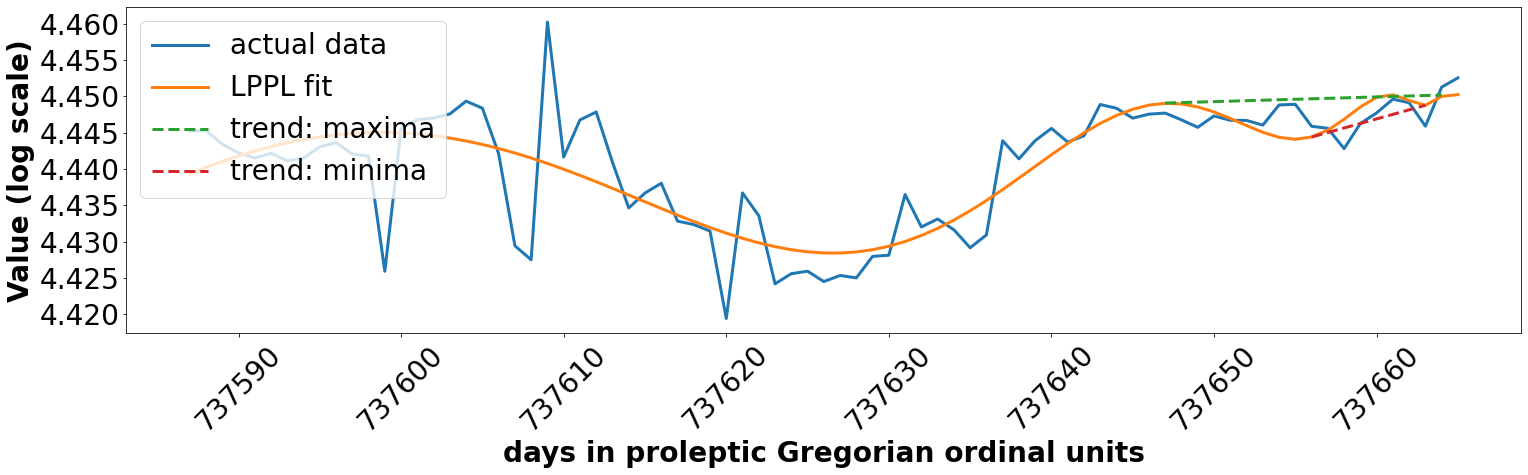} 
        \caption{positive trends}
    \end{subfigure}
    \hfill
    \begin{subfigure}{0.9\textwidth}
        \includegraphics[width=1\linewidth]{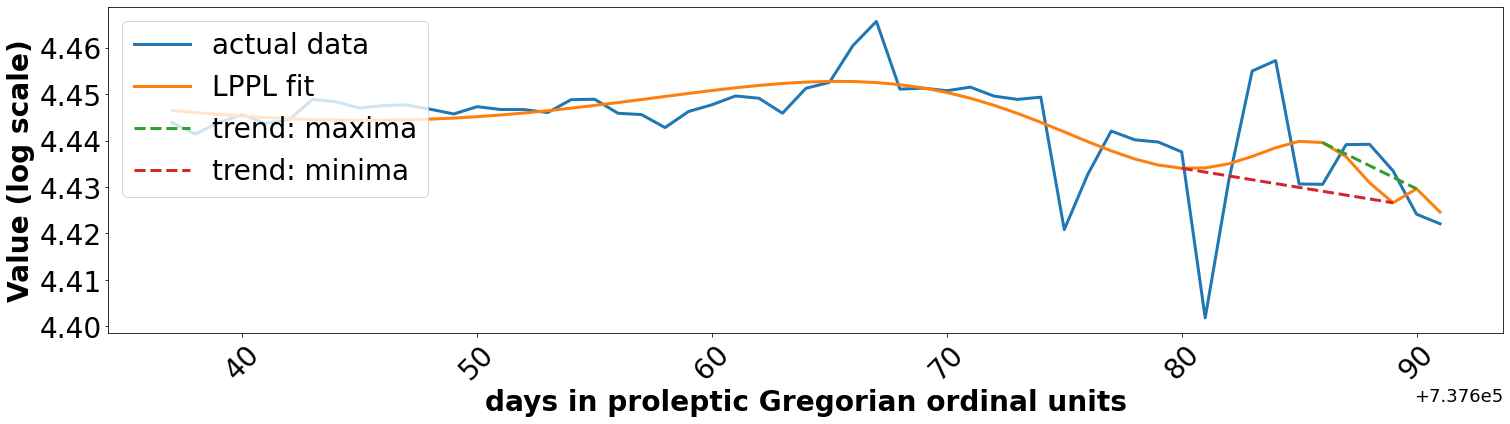}
        \caption{negative trends}
    \end{subfigure}
\caption{Examples of LPPL function (\ref{final_fit_1}) fits for cases with positive trends (a) and with negative trends (b) for selected current times (dates in proleptic Gregorian ordinal units).  Fit parameters calculated for both trend categories:
(a) positive trend: current date: $t_{c}=2020-08-28$, length of the time series: $l_{max}=79$, mean squared error: $mse = 2.6\cdot 10^{-5}$;
(b) negative trend: current date: $t_{c}=2020-09-23$, length of the time series: $l_{max}=55$, mean squared error: $mse = 5.7\cdot 10^{-5}$.}
\label{fig:fit_examples}
\end{figure}

Figure (\ref{fig:fit_examples}) shows 2 examples of the fit function (\ref{final_fit_1}) to data along with calculated criteria for trends determined from maxima and minima of the fitting function satisfying the criteria of Theorem (\ref{bl:theorem1}). 

\begin{figure}[h!]
 \centering
 \includegraphics[width=0.9\textwidth]{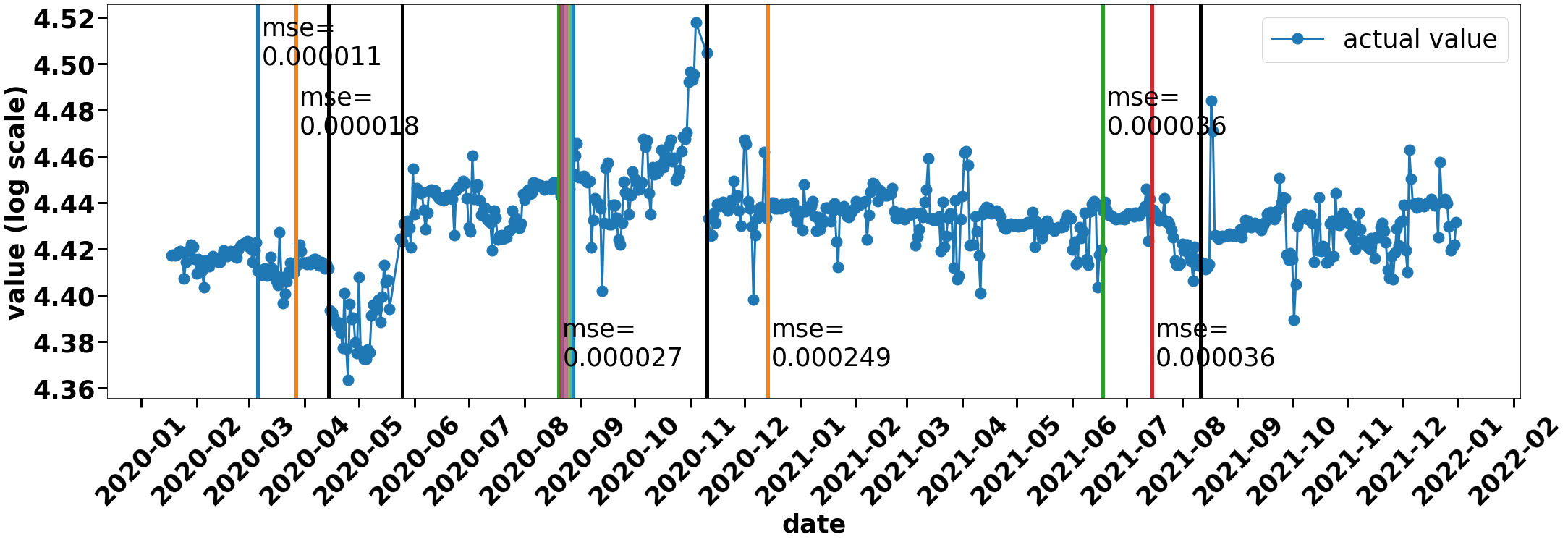}
 \caption{Calculated IB points for the analyzed input time series 
 (marked in figure by a solid line with measurement points). 
 The vertical lines of different colors indicate the diagnosed by the algorithm IB points. 
 For each group of the IB points, the mean value of the matching error ($mse$) of the LPPL function (\ref{final_fit_1}) is annotated. 
 The black vertical lines indicate the dates of compressor repairs, 
 which typically took place a few to several weeks after the algorithm detected the fault (IB points).
 The correlation with the diagnosed breakpoints is clearly visible, 
 except for the prediction determined with the largest error (mse=0.000249) for 2020-12-14.}
\label{fig:detection_examples}
\end{figure}

Figure (\ref{fig:detection_examples}) shows the application of this procedure to the diagnosis of the critical points with additional information about the dates of failure repairs (see the description of Figure (\ref{fig:detection_examples})).   

\noindent
Figure (\ref{fig:detection_examples}) confirms our initial hypothesis very well. It shows:
\begin{enumerate}
\item groups of points with similar fitting errors at least $14$ days before the time of failure identification (repair is usually performed with additional delay due to the compressor operating conditions),
\item dependence of the matching error on the criticality of the failure: the signaling of the grouping of critical points for the beginning of 2020-12-14 is characterized by a large error ($mse=24.9\cdot 10^{-5}$), much larger than for the group of points for which the failure has been confirmed ($mse <= 5.7\cdot 10^{-5}$).
\end{enumerate}

Taking these conclusions into account and comparing the recorded failures and behaviors suggesting problems in compressor operation, classified by experts as insignificant, with the predictions made by the model, it is possible to determine the threshold values of the fitting error ($mse$) and the corresponding categories of predictions:

\begin{definition}\label{bl:definition1}
Classification of calculated initial breakdown points:
\begin{enumerate}
\item {\bf Critical event}: ($mse < 6\cdot 10^{-5}$) severe failure expected, checking of the compressor required and preparation for repair,
\item {\bf Monitoring event}: ($6\cdot 10^{-5}<= mse < 10\cdot 10^{-5}$) distinct possibility of issues, monitoring of compressor behavior required.
\item {\bf Irrelevant event}: ($10\cdot 10^{-5}<= mse$) no significant issues predicted, compressor behaviour can be monitored though.
\end{enumerate}
\end{definition}

As shown, sometimes the algorithm detects a larger number of IB points located very close to each other. Such cases were simplified by choosing a single representation (the first IB point of the group) for each group of signals. This was done by assuming that the signal belongs to a group if its distance from the preceding signal is smaller than or equal to $3$ days.    

\subsection{Determining the time window of predicted failures}
\label{subsection:time_window}

By comparing the failure times predicted by the algorithm with their actual occurrence, a criterion for predicting the time window in which the failure will occur can be also determined.  
One of the parameters for fitting the LPPL function (\ref{final_fit_1}) to the data is the length of the chosen sequence of data preceding the analyzed time point $l_{max}$. 
For the data analyzed, the time window for the occurrence of a predicted failure was defined as:

\begin{definition}\label{bl:definition2}
The predicted time period of a failure occurrence is defined as an interval $\left<n+\frac{l_{max}}{2}, n+90\right>$, 
where $n$ and $l_{max}$ are the indices of the actual time point $x_{n}$ 
and the parameter defining the length of the input time series $t_{inp}$ 
used to find the best fit of the LPPL function (\ref{final_fit_1}) to a given value of 
$t_{inp}\left(x_{n},y_{n}\right)$, respectively. 
\end{definition}
Definition (\ref{bl:definition2}) is based on a knowledge of the dynamics of the device for which the algorithm parameters have been defined. 
For other devices, all parameters should be selected based on the dynamics of their behavior.
Duration of the time window with predicted failure is assumed to be valid for a certain period of time and is up to 90 days.

In a case of compressors, due to the different criticality of failures, some of them may be accepted for a longer period of time (even several months in the case of valve failures) while waiting for a convenient moment for repair.

\begin{figure}[h!]
 \centering
 \includegraphics[width=0.9\textwidth]{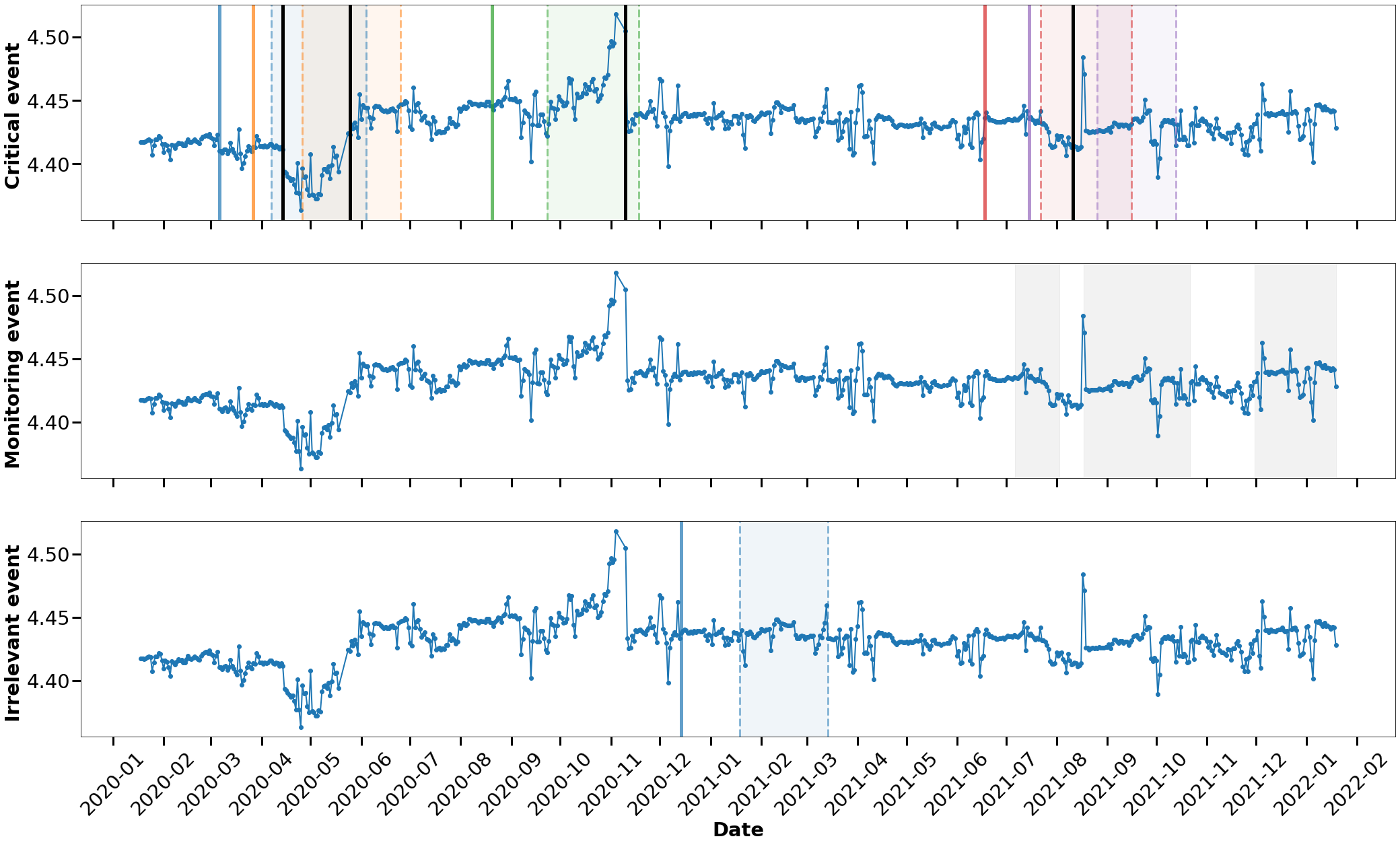}
 \caption{
 Redrawn prediction of failures by the algorithm in comparison with reparation times and problems detected by experts. 
Figure shows representations of groups of discovered IB points (colored and continuous single vertical lines) 
and predicted failure time periods corresponding to them according to Definition (\ref{bl:definition2})
(corresponding colored areas bounded by vertical dashed lines).
The categorization of the criticality of the forecasted problems (Definition (\ref{bl:definition1})) is represented by the splitting into 3 separate figures, each for a separate criticality category. The {\bf Monitoring event} class remains without any marked vertical lines. The algorithm found no prediction for this category. 
The reparation (maintenance) dates are indicated in figures by black vertical lines. Time periods of abnormal compressor behavior requiring monitoring and not qualified for repair by experts are indicated by the gray colour without additional delimiting lines.
The intersection of the areas determined by the algorithm as a period of failure occurrence with periods of abnormal behavior of the compressor can be seen for periods starting from 2021-07-15 in the {\bf Critical event} category.
The raw data of the input time series $t_{inp}$ are indicated by the solid blue line with the measurement points ($o$). }
\label{fig:final_detection_examples}
\end{figure}

Considering: 
\begin{itemize}
\item the classification of alerts specified in Definition (\ref{bl:definition1}),
\item the selection of the representative of the groups of warnings (by selection of the initial signal for the common group of calculated IB points), 
\item Definition (\ref{bl:definition2}) specifying the expected time window in which the failure will occur,
\end{itemize}
the raw results shown in Figure (\ref{fig:detection_examples}) can be redrawn to a new form, as shown in Figure (\ref{fig:final_detection_examples}).
The correlation between the predicted failure times and the actual repair times,
or the time periods when experts detected abnormal compressor behavior is very good for the {\bf Critical event} and {\bf Monitoring event} categories.
Predictions in the {\bf Irrelevant event} category were not confirmed by any repair and diagnosis records.

\subsection{Root cause of predicted failures}
\label{subsection:root_cause}

In the analysis presented here, the input data monitors the change in the opening angle of the suction valves in the compression chamber expressed by the angle of rotation of the crankshaft. 
The trend identified from the determination of the IB points can be used to guess the type of future failure in the compression chamber \cite{Reciprocating_compressors}.
By predicting the trend for times after the IB point, it is possible to try to determine approximately which part will fail - the valve or the piston rod sealing rings.
Thus, when the predicted trend of the suction valve opening angle is decreasing, it is likely that the suction valve or piston rod seal rings are failing. If the trend suggests an increase in angle, this behavior indicates a leak in the discharge valve.

Thus, based on Theorem ({\ref{bl:theorem1}) and a physical interpretation of the behavior of the time series, the algorithm is able to predict not only the time window of failure, but also the group of parts that may fail. 
This provides an opportunity to verify the prediction not only on the basis of event times, but also on the basis of identifying the parts that can fail. 

To better illustrate the additional information regarding the location of the future failure, Figure (\ref{fig:final_detection_examples_root})
shows the same data as Figure (\ref{fig:final_detection_examples}) with additional information about the parts that actually failed, 
the parts in which experts have observed problems and prognosis of failures predicted by the algorithm.
Details are given in the description of Figure (\ref{fig:final_detection_examples_root}). 

For the entire time period analyzed, 2 cases deviating from this diagnosis are visible
\begin{enumerate}
\item
in the category {\bf Monitoring event}, for date $2021-07-15$, there is a disagreement between the predicted failure type {\it suction valve or sealing - leakage} and the diagnosed one {\it indiation of discharge valve leakage}.
\item
the perturbation identified by experts, started on $2021-11-30$ and identified as {\it indiation of discharge valve leakage} was not predicted by the algorithm at all.
\end{enumerate}
\begin{figure}[h!]
 \centering
 \includegraphics[width=0.9\textwidth]{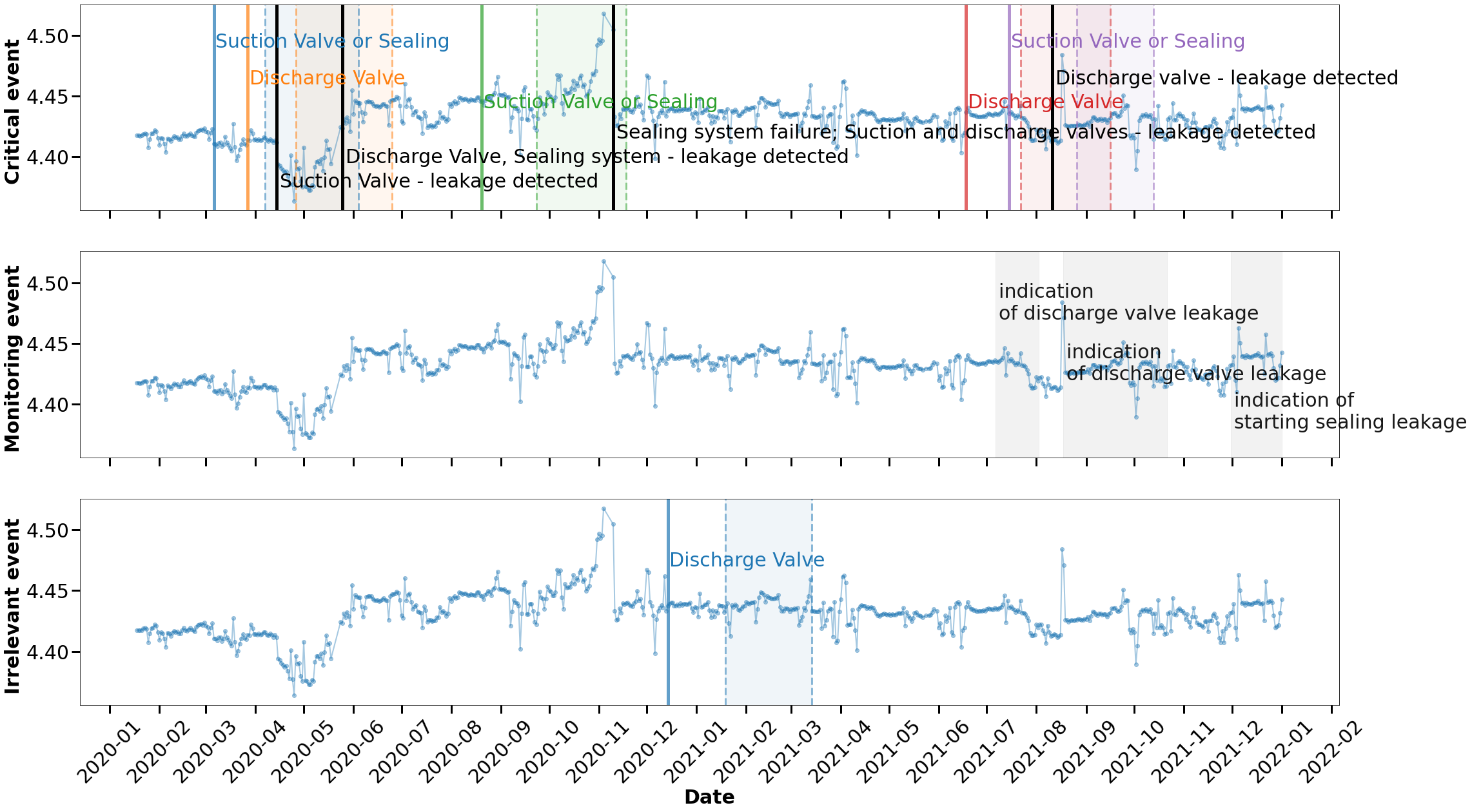}
 \caption{
 The same graphs as in Figure (\ref{fig:final_detection_examples}) with added annotations describing the failures predicted by the algorithm - colored texts (different from gray and black), reasons for compressor reparations - black text and diagnosed abnormal compressor behavior requiring monitoring - gray text. 
 For better visualization, areas of the diagnosed abnormal behaviour of the compressor that require monitoring are displayed in the "Monitored  Event" category. 
 }
\label{fig:final_detection_examples_root}
\end{figure}

\subsection{Comparison with statistical method}
\label{subsection:benchmark} 

The described method generates IB points before the appearance of anomalies associated with a future failure (section \ref{section:fitting}).
This is why the LPPL-based solution has an advantage over solutions based on anomaly detection as a signature of a forthcoming failure.  

It cannot be excluded that the IB points detected by the described method, may be the same as the statistically defined change points in the input trend of the time series. 
For comparison of the results determined by the discussed solution with the trend change points determined statistically, the online changepoint detection method from the work of \cite{changepoint2} was used. 
This solution is available in the form of the Python library \cite{changepoint3} (changepoint\_online). 

I ran a model based on detection of two states of the analyzed data: 
one focusing on the detection of a decreasing trend, and the other focusing on the changes in the increasing trend. 
This way, it is possible to track when the analyzed variable increases and when it decreases. 
This corresponds to the ability to detect the type of future failure, as described in subsection \ref{subsection:root_cause}.

In each iteration of the historical test, two changepoint detectors are run, one focusing on decreasing changes (left detector) and the other on increasing changes (right detector). In the source of the changepoint\_online \cite{changepoint3}, this corresponds to the terminology: left and right detectors, respectively.
The two detectors work independently of each other. 
When either model exceeds a threshold value, it is reinitialized.

A comparison of the results of the two methods at the level of determining IB points and trend change points is shown 
in Figure \ref{fig:comparison_of_methods}. 

\begin{figure}[h!]
 \centering
 \includegraphics[width=0.9\textwidth]{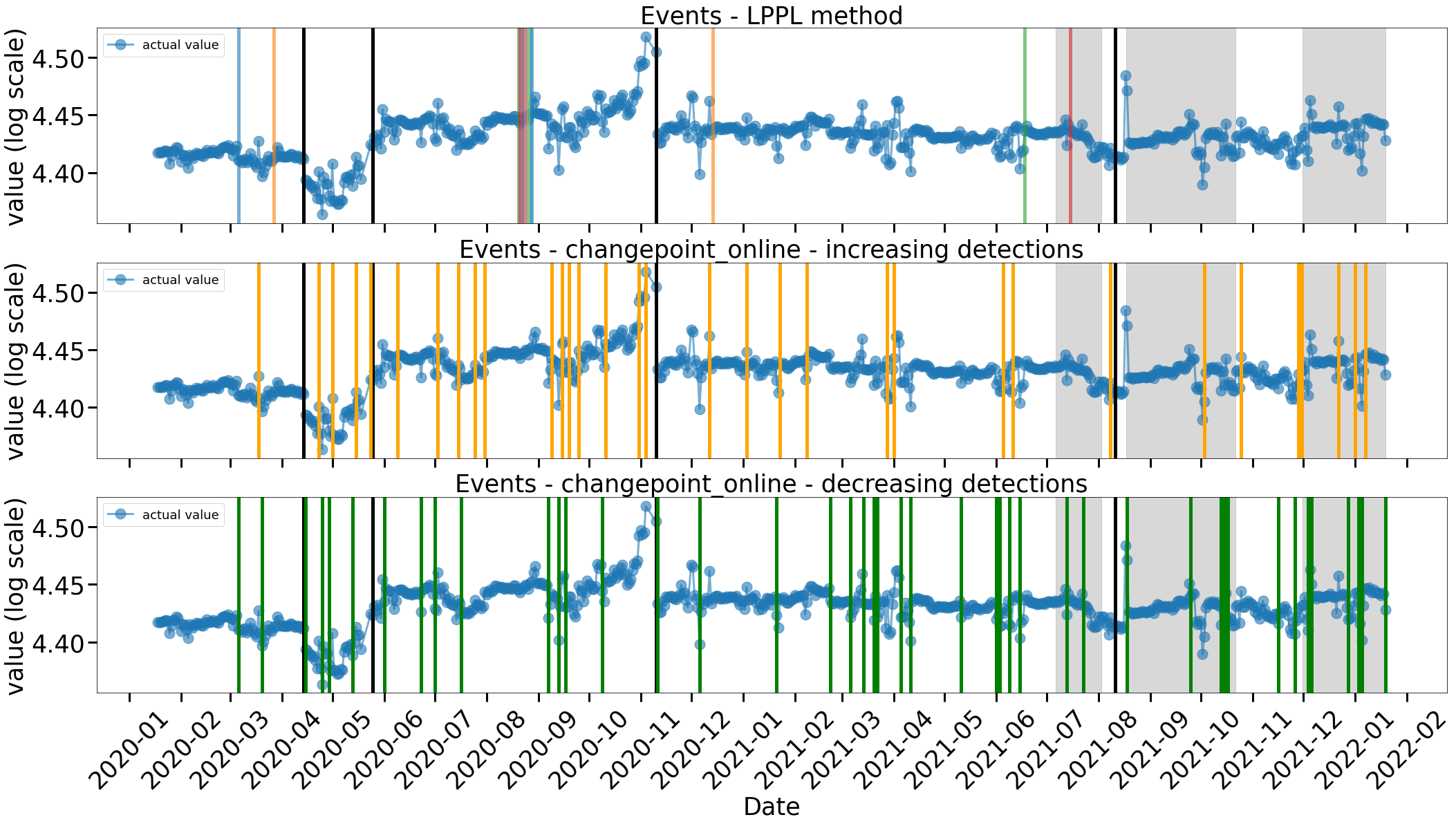}
 \caption{
Comparison of the results of the presented method (LPPL method - top figure) with the statistical method (changepoint\_online - middle and bottom figures) at the level of determining IB points and trend change points. 
The analyzed time series $t_{inp}$ is shown by the solid blue line with the measurement points ($o$). The black vertical lines indicate the compressor repair dates. Time periods when abnormal compressor behavior requiring monitoring was not qualified for repair by experts are indicated by the gray colour without additional delimiting lines.
Top figure: for the LPPL method, the vertical lines in different colors indicate the IB points diagnosed by the algorithm for the LPPL method (as in Figure \ref{fig:detection_examples}). Middle and bottom figure: the found points of upward (middle figure) and downward trend change (bottom figure). Calculations of trend change points are shown for the threshold value of the 75th percentile. 
}
\label{fig:comparison_of_methods}
\end{figure}

As shown in Figure \ref{fig:comparison_of_methods}, the set of IB points determined by the LPPL method (14 alerts) is much smaller than the set of trend change points found by he online statistical method (changepoint\_online) (44 left and 33 right detections) . 
This is consistent with intuition because not every trend change point is a signature of change due to device degradation. 
The selection of all trend change points will create a lot of false signals. 
Therefore, it is necessary to correctly identify the threshold value to be used for selection of the best trend change points, which may suggest future failure. 
The choice of the threshold value should demonstrate the relation between changes in the behavior of the analyzed variable with degradation processes in the monitored device. 
Such a correlation requires a separate analysis and knowledge of the labeled data.
The applied statistical method is unable to predict future failures based solely on determining trend change points in a fully unsupervised manner. 

\section{Discussion} 
\label{section:discussion}

The comparison of information coming from the failure predictions and comparing it with the knowledge from maintenance logs and expert detection of periods of anomalous compressor operation, including the prediction classification, is provided in Table \ref{Table:TableComparison1}.

Major challenges in the field of industrial application of failure prediction, especially in the unsupervised version, is the number of issues corresponding to proper identification of failures and behaviors of monitored devices.
These difficulties mainly stem from:
\begin{itemize}
\item the large variety in the types of failures,
\item the small number of failures compared to the amount of data,
\item the large variety of behaviors leading to the same type of failure,
\item the difference in operating conditions, which can cause ambiguity in labeling accurate data.
\end{itemize}
The problems are very difficult to solve if the methods used to predict failures are based on the analysis of numerical values of input data 
(by calculating similarities, correlations, logic trees, building naural networks, etc.).
Normalization and/or standardization procedures only introduce a common scale to the analyzed data. 

The proposed method introduces a new type of procedure, which is based on the search for common functional behavior (\ref{final_fit_1}).
From the point of view of the numerical values of the analyzed data, the course of the fitted function can be very different for different events - different patterns of the function for different values of fitted parameters.
Even when new data appears with values that did not exist in the past, it is possible to determine the initial breakdown (IB) for a potential event, as the model fits functions to the data.
 
The IB points determined by this method, 
upon which the time window of failure occurrence is predicted in the next step, are the trend change points in the data.

To calculate the key performance indicators (KPIs) of the presented algorithm, the generally known indicators can be used:
$Precision = TP/\left(TP+FP\right)$ and $Recall = TP/\left(TP+FN\right)$, 
where $TP$ - defines {\it True Positive} events, i.e. correctly predicted failures or abnormal behavior in the operation of the compressor, $FP$ - {\it False Positives}, i.e. events predicted by the algorithm that turned out to be false and 
$FN$ - {\it False Negatives}, i.e. failures and instabilities of the compressor that were not predicted.

In the analysis of results (section \ref{section:predictions}), the limits of acceptance of errors of fitting the LPPL function (\ref{final_fit_1}) determining the criticality of the predicted events (Definition \ref{bl:definition1}) were defined on the basis of the results. 
Therefore, in order to calculate the $Precision$ and $Racall$ indicators, all results are taken into account without distinguishing them due to the defined criticality.

Comparing the predicted failure periods, taking into account the dates and predicted types of failures with the dates and descriptions of Maintenance logs or recorded faults (Figures \ref{fig:final_detection_examples} and \ref{fig:final_detection_examples_root} and the table \ref{Table:TableComparison1}), the calculated values of $TP$, $FP$ and $FN$ are as follows:
$TP = 4$, $FP = 2$, $FN = 1$ .
\noindent
Hence $Precision = 0.67$, $Recall = 0.8$. 
Given that, this result takes into account the dating of the predicted failure period along with the prediction of the cause of failure, such an outcome is considered very good.  

In summary, the list of advantages and disadvantages of the presented method is a consequence of a paradigm change in data behavior classification, from the one based on numerical values to the one based on functional similarity.

\noindent
{\bf Advantages of the method}:
\begin{itemize}
\item The proposed model can be applied to very short time series (in our case, the minimum length of the series is only 101 points).
\item There are no problems with data that appears for the first time. In the proposed solution, the part that qualifies certain data as IB is based solely on functional behavior. This universality is due to the renormalization group approach.  
\item The simplicity of the final production solution.
The most difficult part of the algorithm is fitting the function to the data. Since the method does not contain components based on supervised methods, there is no need to monitor their quality. 
\end{itemize}
\noindent   
{\bf Disadvantages of the method}: 
\begin{itemize}
\item The method is applicable to data that describe a physical process that degrades/changes due to perturbations introduced by interacting elements. This is because the method searches for behavior characteristic of phenomena in which phase transitions can be observed. Hence, not all data are appropriate for the described method.
\item Matching the function to the data is based on the proper determination of boundaries of the parameters to be matched (\ref{final_fit_1}). 
This requires individual adjustment of the ranges of change of these parameters and the unit of time in the data to the device being monitored.
\end{itemize}

The method presented is designed to predict such failures, which are a consequence of a continuous physical process that can be monitored by measurements. 
In addition, it is necessary to know the time scales specific to the degradation of the monitored equipment. 
The time unit used by the algorithm should correspond to the time scale units of the degradation process.

\noindent
\begin{sidewaystable}[htbp]
\caption{Detailed comparison of model predictions with maintenance logs entries and expert diagnoses. The "Predictions" section contains a description of the alert: "Alert date" - date of alert occurrence, "mse" - goodness of the fit, "Predicted failure time window" - the predicted time window in which the failure may occur, and "Predicted failure" - the predicted failure diagnosis. 
The "Maintenance logs" section contains columns: "Maintenance date" - repair date and "Diagnosis" - the reason for the failure. 
The "Expert diagnoses" columns contain information about: the beginning date of the anomaly notice - "Start date", the end date of the anomaly - "End date" and the symptom of the anomaly - "Diagnosis".
The "Label" column contains the identification of the predicted failure. The following convention is used:
True Positive (TP) event, with correctly predicted compressor failures or misbehavior, False Positives (FP), an event predicted by the algorithm that turned out to be false, and False Negatives (FN), i.e. compressor failures or instabilities that were not predicted.}
\fontsize{9pt}{9pt}\selectfont
\begin{tabular}{|p{0.05\textwidth}|p{0.08\textwidth}|p{0.08\textwidth}|p{0.11\textwidth}|p{0.09\textwidth}|p{0.11\textwidth}|p{0.1\textwidth}|p{0.08\textwidth}|p{0.08\textwidth}|p{0.09\textwidth}|}
	\hline 
      \multicolumn{5}{|c|}{\textbf{Predictions}} &
      \multicolumn{2}{c|}{\textbf{Maintenance logs}} &
      \multicolumn{3}{c|}{\textbf{Expert diagnoses}} \\
      \hline
    Label & Alert date & mse & Predicted time window of failure & Predicted failure & Maintenance date & Diagnosis & Start date & End date & Diagnosis \\
    \hline
    \hline
    TP & 2020-03-06 & $11\cdot 10^{-5}$ & 2020-04-07 - 2020-06-04 & SV or Sealing & 2020-04-14 & SV: leakage detected & - & - & - \\
    \hline 
    TP & 2020-03-27 & $18\cdot 10^{-5}$ & 2020-04-26 - 2020-06-25 & DV & 2020-05-25 & DV, Sealing system: leakage detected & - & - & - \\
    \hline
    TP & 2020-08-20 & $26\cdot 10^{-5}$ & 2020-09-23 - 2020-11-18 & SV or Sealing & 2020-11-10 & Sealing system failure; SV and DV: leakage detected & - & - & - \\
    \hline
    FP & 2020-12-14 & $249\cdot 10^{-4}$ & 2021-01-19 - 2021-03-14 & DV & - & - & - & - & - \\
    \hline
    TP & 2021-06-18 & $36\cdot 10^{-5}$ & 2021-07-22 - 2021-09-16 & DV &  2021-08-11 & DV - leakage detected & 2021-07-06 & 2021-08-02 & DV leakage \\
    \hline
    FP & 2021-07-15 & $36\cdot 10^{-5}$ & 2021-08-26 - 2021-10-13 & SV or Sealing & - & - & 2021-08-17 & 2021-10-21 & DV leakage \\
    \hline
    FN & - & - & - & - & - & - & 2021-11-30 & 2022-03-02 & starting sealing leakage \\
    \hline
\end{tabular}
\label{Table:TableComparison1}
\end{sidewaystable}

\section{Conclusions}
\label{section:conclusions}

This paper presents the application of a methodology for describing critical behavior in complex systems based on the renormalization group approach in unsupervised predictive maintenance.
The proposed algorithm analyzes the behavior of a complex system based on a time series representing the physical behavior of the system. 
To demonstrate the effectiveness of the algorithm for industrial applications, predictive results are presented for time series describing the thermodynamics of the gas compression process in a monitored reciprocating compressor in one of the compression chambers.

It was shown that failures in the analyzed industrial system can be treated as critical behavior in complex systems.  
Then the symptoms of future failure in the analyzed time series, appear in the form of Log Periodic Power Law structures characteristic of the vicinity of phase transitions in physical systems.
Based on the most generalized scheme for describing the behavior of an analyzed system in the vicinity of phase transitions of the 2nd kind based on the Log Periodic Power Law, a new way of predicting failures in compressor systems is proposed.  

The presented algorithm is based on 3 steps. In the first step, the algorithm determines the IB points in the analyzed time series by means of fitting the LPPL function (equation (\ref{final_fit_1})) using the proposed Theorem (\ref{bl:theorem1}). 
The second step of the proposed method is based on the knowledge of the dynamics of the monitored system, and specifies the time window in which the predicted failure may occur.
In the last step, a criticality classification of the predicted failure is carried out, based on the goodness of fit of the LPPL curve to the data ({\it critical event}, {\it monitoring event}, {\it insignificant event}).

Taking into account the specificity of problem detection in industrial systems (the demand to reduce the number of false alarms and to minimize the number of unpredicted events), it has been demonstrated that it is possible to experimentally determine such an error threshold of fitting the LPPL function to the data that all serious failures can be predicted if the fitting error is smaller than the threshold. 
In addition, it is also possible to define such thresholds for the LPPL curve-fit error to the data, for which the area of occurrence of less critical failures, that do not require rapid intervention, can be defined.

The method can also be applied to predictive IoT analysis of other industrial systems.   

\section*{Acknowledgements}
\label{section:Acknowledgements}
The author thank colleagues from Prognost Systems GmbH for the introduction to the topic of failure detection in compressors 
and for helpful comments and discussions.

\end{document}